\documentclass[final,5p,times,twocolumn,authoryear]{elsarticle}

\usepackage{graphicx}	% Including figure files
\usepackage{amsmath}	% Advanced maths commands
\usepackage{amssymb}	% Extra maths symbols
\usepackage[normalem]{ulem}
\usepackage{array}
\usepackage{comment}

\usepackage[T1]{fontenc}
\usepackage{ae,aecompl}
\usepackage{float}
\usepackage{newtxtext,newtxmath, mathtools}
\usepackage{xcolor}
\usepackage{cancel}

\DeclareMathOperator\erf{erf}

\usepackage{amssymb}
\journal{Astronomy and Computing}

\begin{document}

\begin{frontmatter}



\title{Synthetic Population of Interstellar Objects in the Solar System}


\author[inst1]{Du\v san Mar\v ceta}

\affiliation[inst1]{organization={Department of Astronomy, Faculty of Mathematics, University of Belgrade},%Department and Organization
            addressline={Studentski trg 16}, 
            city={Belgrade},
            postcode={11000},
            country={Serbia}}

\begin{abstract}
The discovery of the first two macroscopic interstellar objects (ISOs) passing through the Solar System has opened entirely new perspectives in planetary science. The exploration of these objects offers a qualitatively new insight into the processes related to the origin, structure and evolution of planetary systems throughout the Galaxy. Knowledge about these phenomena will greatly advance if current and future sky surveys discover more ISOs. On the other hand, the surveys require better characterization of this population in order to improve their discovery algorithms. However, despite their scientific significance, there is still no comprehensive orbital model of ISOs in the Solar System and computationally efficient algorithm for generating their synthetic representations that would respond to these increasing needs. Currently available method for generating synthetic populations cannot fully take into account important phenomena, such as gravitational focusing and the shielding effect of the Sun. On the other hand, this method is also computationally far too demanding to be used for systematic exploration of the ISO population. This paper presents an analytical method for determining the distributions of the orbital elements of ISOs, as well as computationally efficient algorithm for generating their synthetic populations, based on the multivariate inverse transform sampling method. The developed method is several orders of magnitudes more efficient than the available method, depending on the size of the synthetic population. A Python implementation of the method is freely available and can be used to generate synthetic populations of ISOs with user-defined input parameters.
\end{abstract}

\begin{comment}
\begin{highlights}
\item Fading of a script alone does not foster domain-general strategy knowledge
\item Performance of the strategy declines during the fading of a script
\item Monitoring by a peer keeps performance of the strategy up during script fading
\item Performance of a strategy after fading fosters domain-general strategy knowledge
\item Fading and monitoring by a peer combined foster domain-general strategy knowledge
\item The available procedure for generating synthetic populations of Interstellar Objects (ISOs) in the Solar System lacks the computational efficiency necessary for its systematic implementation.
\item A newly developed Probabilistic method for generating synthetic populations of ISOs is orders of magnitude more efficient than the available method.
\item Another advantage of the Probabilistic method is that it treats the effect of gravitational focusing and the shielding effect of the Sun in a complete way ensuring that the synthesized population fully matches the input distribution of interstellar velocities of ISOs.

\end{highlights}
\end{comment}

\begin{keyword}
Planetary systems \sep comets: general \sep minor planets, asteroids: general \sep minor planets, asteroids: individual: 1I/’Oumuamua \sep comets: individual: 2I/Borisov
\end{keyword}

\end{frontmatter}



\section{Introduction}
\label{Introduction}

After the discovery of the first two interstellar visitors, 'Oumuamua \citep{MPC-oumuamua} and Borisov \citep{MPC-borisov} the population of interstellar objects (ISOs) has come into the focus of planetary science and other fields of astronomy and astrophysics. These objects could provide an in-depth insight into the formation, structure and various stages of evolution of planetary systems \citep{Trilling2017}, and may even be the seeds for the formation of planets \citep{2019ApJ...874L..34P, 2022ApJ...924...96M}. Their composition, size, shape and rotational state may be linked to the mechanisms responsible for their ejections in the interstellar space \citep{2018MNRAS.478L..49J, 2018ApJ...852L..15C, 2021A&A...651A..38P, 2022ApJ...935L..31C}, thus offering a unique validation of the models of formation and evolution of the planetary systems. Measurements of production rates of CO$_2$, CO and H$_2$O of future interstellar comets can provide constraints on their primordial C/O ratios, which can be used as an indicator for their formation location within a protostellar disk. In this way, interstellar comets offer a key insight into mechanisms responsible for cometary ejection in exoplanetary systems \citep{2022PSJ.....3..150S}. ISOs could also be laboratories for examination of the interstellar environment which they passed through, the impact of this environment on their sizes, shapes and rotational states \citep{Vavilov2019, Zhou2020, 2021arXiv210904494P}, and galactic weathering processes \citep{2017ApJ...850L..36J, Fitzsimmons2018}. Finally, they could even offer an exclusive opportunity for in situ exploration of these distant worlds, thereby presenting a realistic alternative to the interstellar and cross galactic voyage \citep{2019NatCo..10.5418S, 2021P&SS..19705137M, 2018AJ....155..217S}. A comprehensive review on the ISO population can be found in \citet{2022arXiv220908182J}.

There are many studies that are directly related to the orbital structure of interstellar objects in the Solar System and around other stars, which could benefit from a realistic model of this population and efficient algorithm for generating its synthetic representations. Such a model can be very useful for developing algorithms for determining their orbits \citep[e.g.][]{2021CeMDA.133...41G} and analysing their detection rates \citep{2020MNRAS.498.5386M, 2022PSJ.....3...71H} by the future sky surveys. On the other hand, it is also related to the exploration of dynamical mechanisms for their permanent capture by the planetary systems \citep{2021MNRAS.tmp.3398D} and the Solar System in particular \citep{2021PSJ.....2...53N, 2021PSJ.....2..217N}. ISOs could not only be temporary visitors, but also permanent residents of the Solar System with possible $\sim$8 'Oumuamua analogues inside Jupiter's orbit at any time \citep{2021MNRAS.tmp.3397D}. Moreover, some particular main belt asteroids and Centaurs may be associated with an interstellar origin \citep{2020MNRAS.494.2191N}. However, \citet{2020MNRAS.497L..46M} identify that the Halley-type comets and the Oort cloud are the most likely sources of retrograde co-orbitals and highly inclined Centaurs. 

Numerous models of other Solar System populations have been developed for different purposes. Most refer to the NEO population \citep{2012Icar..217..355G, 2018Icar..312..181G}, but there are also models of other populations, such as Main Belt population \citep{2005AJ....129.2869T}, comets in the outer Solar System \citep{2016AJ....152..103S}, and a comprehensive model of the Solar System which is used to analyse the performance of Pan-STARRS survey and which includes all major populations  \citep{2011PASP..123..423G}.
  
However, despite the high importance of the ISO population, there is still no comprehensive orbital model nor computationally efficient algorithm for generating synthetic representations of this population in the Solar System. The objective of the study presented in this paper is to define the distribution functions of all orbital elements of ISOs in arbitrary volume of space around the Sun, and to develop a procedure for sampling orbits from these distributions, which is computationally efficient enough to enable its systematic application. The synthetic population of ISOs is generated under the assumption that its kinematic distribution is the same as that of stars in the solar neighborhood, as a consequence of dynamical relaxation through the scattering events of giant molecular clouds and dark matter substructure \citep{2018AJ....155..217S, 2022PSJ.....3...71H}. This kinematics is described by the Schwarzschild velocity ellipsoid with respect to the Local Standard of Rest (LSR), with velocity dispersions $\sigma_1$, $\sigma_2$, $\sigma_3$ and the vertex deviation $v_d$ defining the orientation of the ellipsoid with respect to the Galactic reference frame.

This paper is organised as follows.  Section \ref{The effect of gravitational focusing} qualitatively and quantitatively analyzes the effect of gravitational focusing. Section \ref{Dynamical method} describes the currently available methods for generating synthetic population of ISOs along with their limitations and shortcomings. The initial version of this method does not take into account the effect of gravitational focusing outside the sphere in which the synthetic population is generated. Since this method is based on the propagation of the equation of motion, i.e. hyperbolic Kepler equation, it will hereafter be referred to as the \emph{Dynamical method (without focusing)}. Subsection \ref{Dynamical method - modification} describes a modification of this method in order to take into account the effect of gravitational focusing outside the sphere in which the synthetic population is generated. This modification will hereafter be referred to as \emph{Dynamical method (with focusing)}.
Section \ref{Probabilistic method} describes a newly developed method to overcome the drawbacks of the existing methods. Since this method is based on the sampling from the probability distributions of the orbital elements, it will be hereafter referred to as the \emph{Probabilistic method}. Section \ref{Results and discussion} summarizes the results of the study and it is divided into three subsections: Subsection \ref{Results 1} which compares synthetic populations of ISOs with different distributions of interstellar velocities, Subsection \ref{Results 2} which analyzes accuracy and computational efficiency of the newly developed method and its comparison with the existing ones, and Subsection \ref{Results 3} which evaluates limitations of the developed method. Section \ref{Summary and Conclusions} summarises the most important conclusions of the work.

\section{The effect of gravitational focusing}
\label{The effect of gravitational focusing}

Considered homogeneous in interstellar space, a structure of the ISO population
in the Solar System is disturbed by the influence of the Sun's gravity, and to significantly lesser extent by the gravity of the other Solar System objects. As shown later, planets can significantly affect the orbits of individual objects, but they cannot disturb the structure of the ISO population. Therefore, when generating synthetic population of ISOs using the \emph{Probabilistic method}, only the gravity of the Sun is taken into account.

Semi-major axis ($a$), eccentricity ($e$) and perihelion distance ($q$) of an ISO are defined by its interstellar velocity ($v_{\infty}$) and impact parameter ($B$) \citep[e.g.][]{kemble2006}.

\begin{equation}
\begin{aligned}
&a=-\frac{\mu}{v_{\infty}^2}\\
&q=a+\sqrt{a^2+B^2}\\
&e=\sqrt{1+\frac{B^2}{a^2}}
\end{aligned},
\label{eq:a,q,e}
\end{equation}
where $\mu$ is the standard gravitational parameter of the Sun.

\begin{figure}[h]
	\includegraphics[width=\columnwidth]{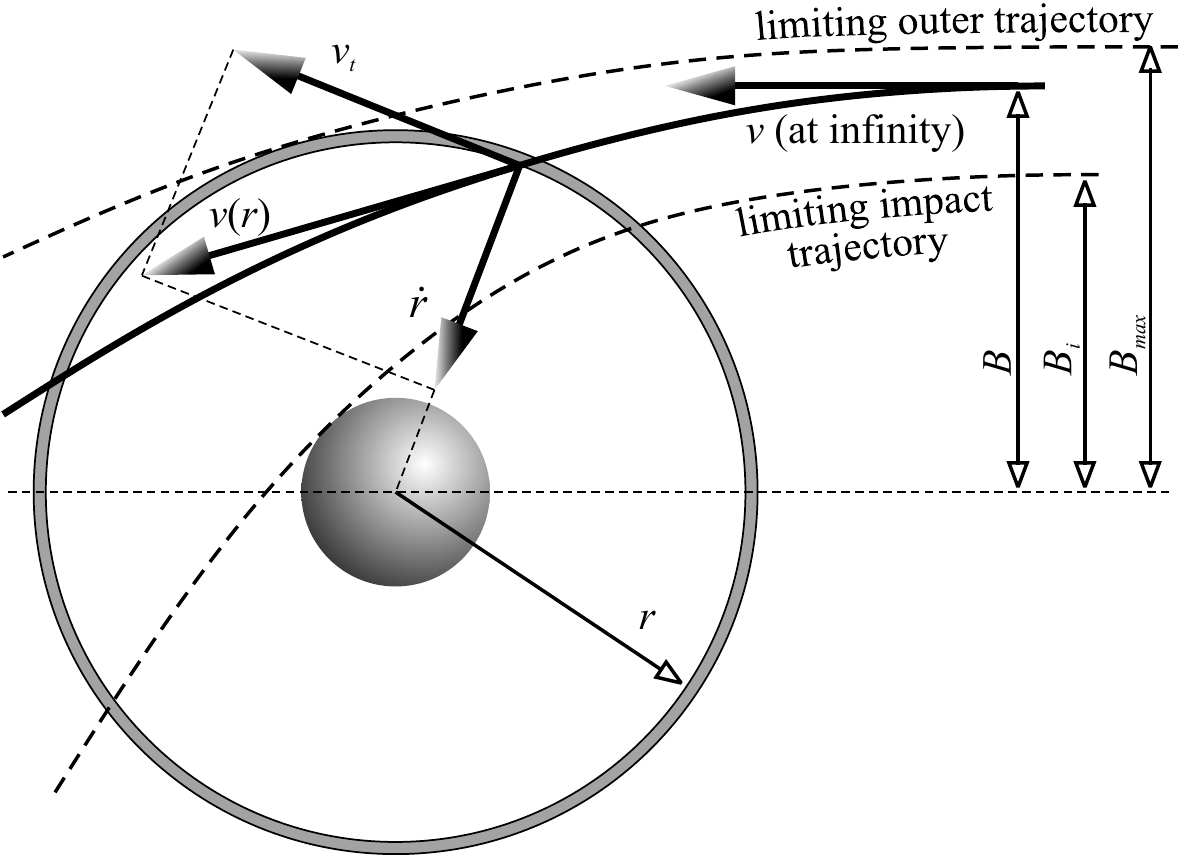}
    \caption{Schematic illustration of incoming trajectory of an ISO  crossing an infinitesimal heliocentric spherical shell and characteristic boundary trajectories.}
    \label{fig:limit}
\end{figure}

The interstellar velocity $v_{\infty}$ is defined as the magnitude of the heliocentric velocity vector when the object is at a very large heliocentric distance, so that the effect of the Sun's gravity can be neglected (the so-called hyperbolic excess velocity). On the other hand, the impact parameter $B$ is the distance of this velocity vector from the Sun (see Figs. ~\ref{fig:limit} and ~\ref{fig:geometry}).

In a qualitative sense, the effect of gravitational focusing works as follows. The Sun's gravity curves the orbits of slower objects more than the orbits of faster ones. According to Eqs.  ~\ref{eq:a,q,e}, an object with a lower interstellar velocity (energy) also has a smaller semi-major axis (larger by absolute value). Accordingly, this object has a smaller perihelion distance and eccentricity for the same impact parameter. This means that a decrease in interstellar velocity increases the effective cross-section that objects have to traverse at infinity in order to reach a certain heliocentric distance. The situation is similar with the impact parameters. For the same interstellar velocity, an object with smaller $B$ has smaller $q$ and $e$ compared to an object with larger $B$. Schematic illustration of this effect is shown in Fig. ~\ref{fig:limit}.

As a consequence of this effect, through a heliocentric spherical shell of arbitrary radius $r$ pass objects whose impact parameters are smaller than or equal to $r$ (as it would be the case without the effect of gravitational focusing), but also objects with larger impact parameters whose orbits are bent enough to reach heliocentric distance $r$. The maximum impact parameter that allows an object to reach an arbitrary heliocentric distance can be easily determined from the conservation of energy and angular momentum,  
\begin{equation}
\begin{aligned}
&\frac{1}{2}\left(v_t^2+ v_r^2\right)-\frac{\mu} {r}=\frac{1}{2}v_{\infty}^2, \\
&r \cdot v_t=B \cdot v_{\infty},
\label{eq:conservations}
\end{aligned}
\end{equation}
where $v_t$ is tangential and $v_r$ is radial component of an ISO's velocity vector at arbitrary heliocentric distance $r$.

The expression for radial velocity at an arbitrary heliocentric distance $r$ can be derived from Eqs. ~\ref{eq:conservations},

\begin{equation}
v_r=\frac{1}{r} \sqrt{v_{\infty}^2r^2+2\mu r-B^2v_{\infty}^2}.
\label{eq:r_dot}
\end{equation}

For a given $v_{\infty}$, the limiting impact parameters $B_1$ and $B_2$ (see Fig. ~\ref{fig:limit}) can be determined from Eq. ~\ref{eq:r_dot} by setting $v_r=0$ for the bounding outer trajectory, and $r=R$ ($R$ is solar radius) for the bounding impact trajectory, giving

\begin{equation}
\begin{aligned}
&B_1=R\sqrt{1+\frac{2\mu}{v^2R}}, \\
&B_2=r\sqrt{1+\frac{2\mu}{v^2r}}.
\label{eq:B_lim}
\end{aligned}
\end{equation}

From the expression for the maximum impact parameter in Eq. ~\ref{eq:B_lim}, a ratio between effective and geometrical cross-sections can be obtained in the form

\begin{equation}
\left(\frac{B_2}{r}\right)^2=1+\frac{v_{esc}^2}{v^2},
\label{eq:cross-section}
\end{equation}
where $v_{esc}=\sqrt{2\mu/r}$ is the escape velocity at heliocentric distance $r$. As already pointed out, Eq. ~\ref{eq:cross-section} shows that the effective cross-section, which any object has to pass at infinity in order to reach heliocentric distance $r$, increases if the interstellar velocity decreases. This means that at any heliocentric distance, the increase in number-density will be greater for slower objects in the population because these objects have larger effective cross-sections, and thus can reach that heliocentric distance from a larger volume of space compared to faster objects. According to Eq. ~\ref{eq:cross-section}, if the heliocentric distance decreases, the ratio between the effective and geometric cross-section increases, which is why the effect of gravitational focusing is more pronounced at small heliocentric distances. However, due to the physical dimensions of the Sun, at some point, a so-called shielding effect of the Sun becomes dominant over the effect of gravitational focusing, which leads to a decrease in number-density of the population with a further decrease in heliocentric distance. This effect will be elaborated in more detail later. 
   
It follows from the above analysis that the gravity of the Sun affects the ISO population by increasing their number-density relative to interstellar space. This increase depends on heliocentric distance and interstellar velocities of ISOs. The resulting influence on the distributions of the orbital elements of ISOs around the Sun is the most important problem to be solved in order to generate their synthetic population. 

\section{Dynamical method for generating synthetic population of ISO\lowercase{s}}
\label{Dynamical method}

Previous analyzes involving synthetic populations of ISOs \citep[e.g.][]{Engelhardt2017, 2020MNRAS.498.5386M} used a technique developed by \citet{2011PASP..123..423G}. This technique is based on propagating an initially homogeneous and isotropic population over a sufficiently long time so that the resulting population represents a steady-state population altered by the effect of gravitational focusing in some predefined sphere surrounding the Sun. To resolve the effect of gravitational focusing on the initially homogeneous population, this method consists of the following steps:
\begin{enumerate}
    \item Defining a sphere around the Sun where the ISO population will be generated. This sphere is called the \emph{model sphere}. It is assumed that the population is initially homogeneous in this sphere and its kinematics follow a distribution assumed to represent realistic kinematics of ISOs in interstellar space.
    \item Calculating the time required for this sphere to be completely emptied of all initial objects. This time is called the \emph{initialization time}. This ensures that, at the end of the propagation, this sphere is completely populated by objects that were initially outside this sphere, where the gravitational focusing is considered to be negligible, so the assumed homogeneous population is very similar to the real population of ISOs.
    \item Calculating the maximum heliocentric distance from which the fastest object from the population can reach the \emph{model sphere} during the \emph{initialization time}. A sphere with this radius is called the \emph{initialization sphere}. Objects outside this sphere cannot reach the \emph{model sphere} during the \emph{initialization time}, so these objects are not taken into account. 
    \item The initial population is generated inside the \emph{initialization sphere} and propagated for the \emph{initialization time} by solving the hyperbolic Kepler equation. All objects located inside the \emph{model sphere} at the end of the \emph{initialization time} make up the synthetic population of ISOs inside this sphere.

\end{enumerate}

This procedure results in the \emph{model sphere} being populated only with objects that were initially outside this sphere, where it is assumed that the gravitational focusing is negligibly small. Although this procedure allows the generation of ISO population that is very similar to the real population, it still has two major drawbacks. The first is that it requires setting a certain heliocentric distance beyond which the gravitational focusing can be completely neglected. As discussed earlier, Eq. ~\ref{eq:cross-section} which can be considered a simple quantification of the gravitational focusing shows that this effect is stronger for slower objects than for faster ones. This means that setting any heliocentric distance beyond which this effect is neglected will lead to a certain underestimation of slower objects in the final synthetic population. Although this underestimation is small, it may be significant because slower objects may more likely be discovered by the sky surveys or captured in the Solar System by some of the dynamical mechanisms \citep{2021PSJ.....2...53N, 2021MNRAS.tmp.3397D}.

The other drawback of this procedure is that it is very computationally demanding. To illustrate the magnitude of the computational load required by this method, a brief analysis is given for the generation of population inside the \emph{model sphere} with radius of 100 $au$ and with a range of initial velocities between 1 $km/s$ and 100 $km/s$. The minimum velocity defines the \emph{initialization time}, required for all objects to leave the \emph{model sphere}, which in this case is about 205 years. This time, together with the maximum predefined velocity, defines the radius of the \emph{initialization sphere}, which in this case is about 4500 au. This results in an enormous initialization volume of about 382 billion $au^3$. This volume is almost 5 orders of magnitude larger than the volume of the \emph{model sphere}, resulting in a similar ratio of the numbers of initial objects and final objects that make up the synthetic population of ISOs. Number-density of ISOs is very unconstrained, with estimations ranging from $10^{-9}\, au^{-3} \left( \sim 10^7 \, pc^{-3}\right)$ \citep{2009ApJ...704..733M} to $\sim 10^{-1}\, au^{-3} \left( \sim 10^{15} \, pc^{-3}\right)$ \citep{2017RNAAS...1...43L, 2017ApJ...850L..38T, 2017ApJ...850L..36J, 2018ApJ...855L..10D, 2021ApJ...922...39L}. However, even for the most conservative assumptions about number-density and reasonable size-frequency distributions, it is quite easy to achieve number-density of 1 object per $au^3$ for 10 meter-sized objects already, which could for instance be detected by the Vera Rubin Observatory's Legacy Survey of Space and Time \citep{2021RNAAS...5..143S}. This results in a very large number of objects whose states have to be determined at the beginning and end of the \emph{initialization time} in order to select those which are within the \emph{model sphere} at that moment. This procedure requires about 2000 hours on 6-Core 3.0GHz CPU. Increasing the range of initial velocities, size of the model sphere and/or number-density results in a further increase in computational load.

\subsection{Modification of the Dynamical method}
\label{Dynamical method - modification}
The described procedure ignores gravitational focusing outside the model sphere. If the \emph{model sphere} is small, this leads to significant underestimation of the number of objects in the resulting population, especially in the lower part of the velocity distribution. To overcome this problem, the initial population outside the model sphere can be modified according to the method applied in \citet{2018AJ....155..217S} and \citet{2022PSJ.....3...71H}. The \emph{initialization sphere} is divided into spherical shells, and the velocity distribution is divided into narrow bands. For each band $\Delta v=\left(\Delta v_1, \Delta v_2, \Delta v_3\right)$, the spatial number-density of ISOs is given by

\begin{equation}
n\left(\Delta v \right)=\frac{n}{8}\prod_{i = 1}^{3}\erf \left(\frac{v_i}{\sqrt{2}\sigma_i}\right)\left|{}^{v_i + 
 \Delta v_i}_{v_i -
 \Delta v_i} \right|.
\end{equation}

The increase in number-density for objects falling within these velocity ranges is calculated in each spherical shell according to

\begin{equation}
\xi=\sqrt{1+\left(\frac{v_{esc}}{v_{\infty}}\right)^2}.
\end{equation}
where $v_{esc}=\sqrt{2\mu/r}$ is the solar escape velocity in the middle of each spherical shell. After the initial positions of ISOs are generated, to account for the solar acceleration, each velocity vector is multiplied by a factor $\sqrt{v_{\infty}^2+v_{esc}^2}/v_{\infty}$.

This modification makes it possible to largely overcome one of the main drawbacks of the previously described procedure, but its result still deviates slightly from the real population. Although it adjusts the speed of objects located at arbitrary heliocentric distances due to the solar acceleration, the directions of the velocity vectors are assumed to be the same as at infinity. Since solar acceleration also bends the orbits of ISOs, the velocity vectors at an arbitrary heliocentric distance point closer to the Sun compared to the corresponding interstellar velocity vectors. Neglecting this the objects are given a slightly larger angular momentum, and this results in a comparably larger eccentricity and perihelion distance. This effect is very small and noticeable only for very small model spheres, as shown in section ~\ref{Results 2}.

\section{\emph{Probabilistic method} for generating synthetic population of the ISO\lowercase{s}}
\label{Probabilistic method}

A completely different approach can be applied in order to overcome the limitations of the \emph{Dynamical methods}. The idea is to derive probability functions of the orbital elements of ISOs and to use them to directly sample their orbits using the inverse transform sampling method \citep[e.g.][]{Devroye1986}. This approach has two advantages over the \emph{Dynamical methods}. Firstly, it allows the treatment of the effect of gravitational focusing without the restriction used in the \emph{Dynamical methods}. Secondly, it allows direct sampling of orbits only for the objects that make up the synthetic population, without the need to treat an enormous number of initial objects of which only a tiny part makes the final population, as is the case with the \emph{Dynamical methods}. For the sake of notation, boldface symbols with subscript \emph{s} denote generated sets of the orbital elements (e.g. $\boldsymbol{q_s}$ denotes the set of perihelion distances), while the same regular type face symbols denote arbitrary elements from the corresponding set (e.g. $q_s$ denotes an arbitrary element from the set of perihelion distances $\boldsymbol{q_s}$).

\subsection{Deriving probability distribution of orbital parameters of ISOs}

\begin{figure}[h]
	\includegraphics[width=\columnwidth]{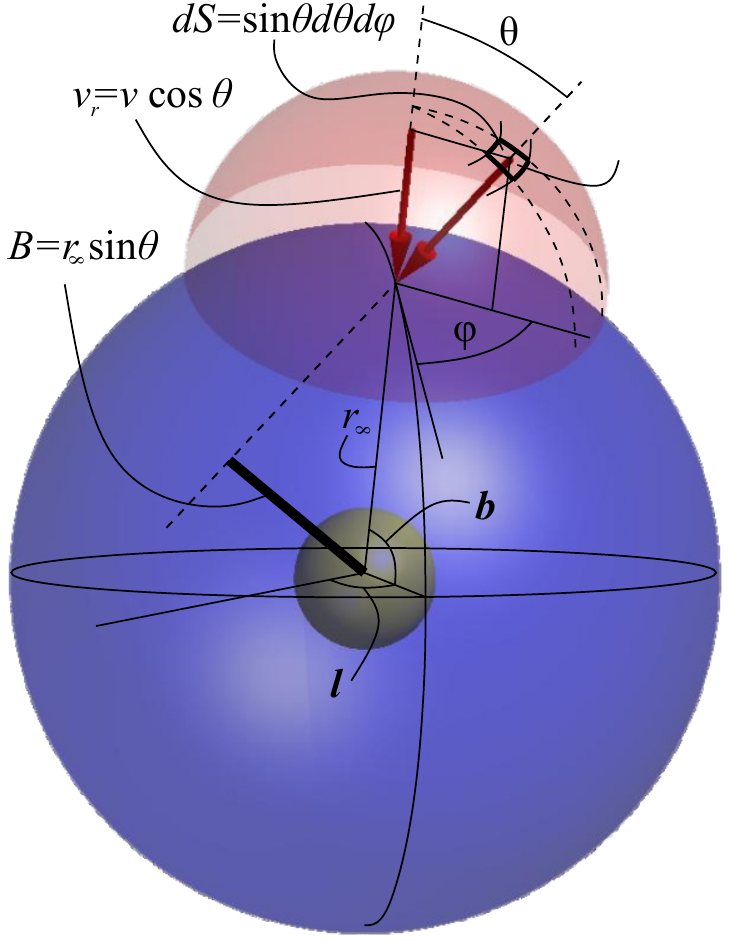}
    \caption{Geometrical constellation of an ISO approaching the Sun from infinity.}
    \label{fig:geometry}
\end{figure}

Early works by \citet{Beard1959} and \citet{Hale1964} analysed the increased concentration of interplanetary meteoroids compared to interplanetary space due to gravitational focusing by the Earth. \citet{Colombo1966} and \citet{Singer1961} derived analytical expressions to describe the increased flux of the interplanetary meteoroids at arbitrary geocentric distance. These expressions can be used as a basis for deriving the probability density function of the orbital elements of ISOs in the Solar System. The geometrical constellation of an ISO approaching the Sun from infinity is illustrated in Fig. ~\ref{fig:geometry}.

The total number of objects per unit time that enter a sphere of radius $r_{\infty}$ around the Sun is

\begin{equation}
\begin{aligned}
&F=\\
 &4\pi r_{\infty}^2 \int \limits_{v_{\infty\left(min\right)}}^{v_{\infty\left(max\right)}} \int \limits_{0}^{2\pi} \int \limits_{-\frac{\pi}{2}}^{\frac{\pi}{2}} \int \limits_{0}^{\frac{\pi}{2}} \int \limits_{0}^{2\pi} n v_{\infty} \cos{\theta}\sin{\theta} \,dv_{\infty} \,dl \,db \,d\theta \,d\varphi,
\label{eq:F}
\end{aligned}
\end{equation}
where $n$ is the total interstellar number-density of objects, which generally, for anisotropic population, depends on interstellar velocity vector defined by $v_{\infty}$,  $l$, $b$, $\theta$ and $\varphi$, where $l$ and $b$ are galactic longitude and latitude, respectively, and angles $\theta$ and $\varphi$ are defined in Fig. ~\ref{fig:geometry}.

When radius $r_{\infty}$ approaches infinity, velocity vector becomes hyperbolic excess velocity vector and the impact parameter $B$ can be defined as $B=r_{\infty}\sin \theta$, (see. Fig. ~\ref{fig:geometry}). On the other hand, angle $\theta$ approaches zero for all objects that can reach an arbitrary heliocentric distance. Consequently, at infinity, number-density of objects capable of reaching arbitrary heliocentric distance is not a function of $\theta$ and $\varphi$, but only of $v_{\infty}$, $l$ and $b$. Integrating with respect to $\theta$ and $\varphi$ leads to 

\begin{equation}
n_{v_{\infty},l,b}=4 \pi n_0 p_{v_{\infty},l,b},
\label{eq:n_v,l,b}
\end{equation}

where $n_0$ is the total interstellar number-density, and $p_{v_{\infty},l,b}$ is the distribution of ISOs with respect to magnitude and direction of interstellar velocity vector. Now Eq. ~\ref{eq:F} can be reformulated into

\begin{equation}
\begin{aligned}
&F=\\
 &n_0  \int \limits_{v_{\infty\left(min\right)}}^{v_{\infty\left(max\right)}} \int \limits_{0}^{\infty}\int \limits_{0}^{2\pi}\int \limits_{-\frac{\pi}{2}}^{\frac{\pi}{2}} \int \limits_{0}^{2\pi} p_{v_{\infty}, l, b} v_{\infty} B \,dv_{\infty} \,dB \,dl \,db \,d\varphi.
\label{eq:F_2}
\end{aligned}
\end{equation}

Eq. ~\ref{eq:F_2} presents the total (integral) flux of ISOs. Differentiation of Eq. ~\ref{eq:F_2} with respect to $v_{\infty}$, $B$, $l$, $b$ and $\varphi$, gives the differential flux that depends on these five variables, in the form

\begin{equation}
F_{v_{\infty}, B, l, b,\varphi}= n_0 p_{v_{\infty}, l, b} v_{\infty} B.
\label{eq:F_3}
\end{equation}

As one can see, this function does not depend on $\varphi$. This means that this angle can be drawn from a uniform distribution from $\left(0, 2\pi \right)$, as it was done in \citet{2018AJ....155..217S} and \citet{2022PSJ.....3...71H}.

Out of all objects coming from infinity, only a fraction will penetrate deep enough to enter a sphere of arbitrary radius $r$. Depending on their $v_{\infty}$ and $B$, some of these objects will cross an infinitesimally thin shell of radius $r$ twice, and some of these objects will impact the Sun, so they will do that only once (see Fig. ~\ref{fig:limit}). With this in mind, at an arbitrary heliocentric distance $r$, the former objects will contribute to number-density twice as large as the latter ones.

Due to conservation of the particles, the total number-density of objects in a spherical shell with arbitrary radius $r$ (i.e. with heliocentric distances between $r$ and $r+dr$), with interstellar velocities between $v_{\infty}$ and $v_{\infty}+dv_{\infty}$, impact parameters between $B$ and $B+dB$, velocity direction angle between $\varphi$ and $\varphi + d \varphi$, coming from the part of the sky bounded by $l$, $l+dl$, $b$ and $b+db$, can be obtained by dividing the flux defined by Eq. ~\ref{eq:F_3} by the area of the shell ($4r^2\pi$) and radial velocity of objects defined by Eq. ~\ref{eq:r_dot}, giving

\begin{equation}
\begin{aligned}
&n_{r, v_{\infty}, B, l, b, \varphi}=n_0 p_{r, v_{\infty}, B, l, b, \varphi}=n_0 p_6=\\
&\begin{cases}
\displaystyle
n_0\frac{v_{\infty}p_{v_{\infty}, l, b} }{4\pi r}\frac{B}{\sqrt{v_{\infty}^2r^2+2\mu r-B^2v_{\infty}^2}}, & \text{$B \leq B_1$}\\\\\\
\displaystyle
n_0\frac{v_{\infty}p_{v_{\infty}, l, b} }{2\pi r}\frac{B}{\sqrt{v_{\infty}^2r^2+2\mu r-B^2v_{\infty}^2}}, & \text{$B_1< B \leq B_2$}
\end{cases}
\end{aligned},
\label{eq:n_r,v,B,l,b,phi}
\end{equation}
where $B_1$ and $B_2$ are the critical impact parameters defined by Eq. ~\ref{eq:B_lim} and illustrated in Fig. ~\ref{fig:limit}. Function $p_{r, v_{\infty}, B, l, b,\varphi}$, defined in Eq. ~\ref{eq:n_r,v,B,l,b,phi} is a joint probability density functions of 6 parameters which uniquely define orbits of ISOs. This function will be used to sample their orbits in arbitrary volume of space around the Sun and for arbitrary distribution of their interstellar velocity vectors. To simplify notation, this function will hereafter be refereed to as $p_6$. 

As mentioned earlier, kinematics of ISOs in interstellar space can be described by the Schwarzschild distribution. To transform Schwarzschild velocity ellipsoid in the Galactic $\left(U, V, W\right)$ system to spherical $\left(v_{\infty}, l, b\right)$ system, it is multiplied by the Jacobian of the transformation,

\begin{equation}
p_{v_{\infty}, l, b}=f\left(U,V,W\right)\begin{vmatrix}
\frac{\partial U}{\partial v_{\infty}} & \frac{\partial U}{\partial l} & \frac{\partial U}{\partial b}\\\\
\frac{\partial V}{\partial v_{\infty}} & \frac{\partial V}{\partial l} & \frac{\partial V}{\partial b}\\\\
\frac{\partial W}{\partial v_{\infty}} & \frac{\partial W}{\partial l} & \frac{\partial W}{\partial b}
\end{vmatrix}.
\label{eq:jacobian}
\end{equation}

Taking into account that 

\begin{equation}
\begin{bmatrix}
U\\\\ V\\\\ W
\end{bmatrix}=\begin{bmatrix}
v_{\infty} \cos b \cos l\\\\ v_{\infty} \cos b \sin l\\\\ v_{\infty} \sin b
\end{bmatrix},
\label{eq:u,v,w}
\end{equation}
Eq. ~\ref{eq:jacobian} becomes

\begin{equation}
p_{v_{\infty}, l, b}=f\left(U,V,W \right)v_{\infty}^2\cos b.
\label{eq:p_v,l,b1}
\end{equation}

For the purpose of demonstrating the \emph{Probabilistic method}, three different Schwarzschild distributions were used. Parameters of these distributions with respect to the LSR are given in Table ~\ref{table:2} \citep{2018AJ....155..217S, 2022PSJ.....3...71H, Binney1998}.  Solar motion with respect to the LSR is defined by the velocity components - $U_\odot=10$ km/s, $V_\odot=11$ km/s, $W_\odot=7$ km/s \citep{2016ARA&A..54..529B}. However, the \emph{Probabilistic method} can be equally applied to any other distribution of interstellar velocities.

\begin{center}
\begin{table}
\begin{tabular}{ |m{2.5cm} c c c c| }
\hline
Stellar type & $\sigma_R$ & $\sigma_{\phi}$ & $\sigma_Z$ &  Vertex \\ {}& {}&{} &{} & deviation $(^\circ)$ \\\hline\hline
M & 31 & 23 & 16 & 7\\\hline
G & 26 & 18 & 15 & 12\\\hline
O$/$B & 12 & 11 & 9 & 36\\\hline

\end{tabular}
\caption{Parameters in the Schwarzschild Distributions for Different Stellar Types.}
\label{table:2}
\end{table}
\end{center}

\subsection{Generating samples of orbital parameters of ISOs}

As mentioned earlier, parameters $r$, $v_{\infty}$, $B$, $l$, $b$ and $\varphi$ uniquely define orbits of ISOs. This makes function $p_6$ (Eq. ~\ref{eq:n_r,v,B,l,b,phi}) sufficient to generate a synthetic population of ISOs in arbitrary volume of space around the Sun. According to the multivariate inverse transform sampling method, which is used for generating synthetic orbits of ISOs, a parameter is sampled by inverting a joint probability function which is cumulative with respect to that parameter, conditional with respect to parameters already determined, and marginal with respect to parameters yet to be determined. In order to distinguish the four types of probability functions that appear during the application of this method, a special notation is used. The joint probability density function of some parameters is denoted by putting those parameters in the subscript (e.g. $p_{r,v_{\infty},B, l, b, \varphi}$ is a joint probability density function of $r$, $v_{\infty}$, $B$, $l$, $b$ and $\varphi$). The distribution function that is cumulative with respect to some parameter is denoted by adding \emph{cdf} to that parameter in the function name (e.g. $p_{r,v_{\infty}, B_{cdf}, l, b,\varphi_{cdf}}$ is a probability density function with respect to $r$, $v_{\infty}$, $l$ and $b$, but cumulative with respect to $B$ and $\varphi$). In the case of a marginal probability function with respect to some parameter, that parameter is omitted from the subscript in the function name (e.g. $p_{r,v_{\infty}}$ is the probability density function with respect to $r$ and $v_{\infty}$ that is marginal with respect to all other parameters, which means that it includes objects with all possible values of these parameters). Finally, when the distribution is conditional with respect to some parameters, those parameters are put in the brackets (e.g. $p_{v_{\infty cdf}} \left(r_0, B_0\right)$ is a function which is cumulative with respect to $v_{\infty}$ , corresponds to some specific heliocentric distance $r_0$ and impact parameter $B_0$, and marginal with respect to other parameters).

The resulting generated sample is independent of the sampling order of the parameters. However, to sample ISO orbits in an arbitrary sphere around the Sun, the total number of objects in that sphere must first be determined. The integration of function $n_{r, v, \varphi, B, l, b}$ (Eq. ~\ref{eq:n_r,v,B,l,b,phi}) over the whole domain of all parameters except $r$ gives the variation of number-density of ISOs with heliocentric distance, defined by

\begin{equation}
\begin{aligned}
&n_r= n_0 p_r = n_0 \int \limits_{v_{\infty\left(min\right)}}^{v_{\infty\left(max\right)}}  \int \limits_{0}^{B_2}\int \limits_{0}^{2\pi}\int \limits_{-\frac{\pi}{2}}^{\frac{\pi}{2}} \int \limits_{0}^{2\pi}\\ & p_6 \,dv_{\infty} \,dB \,dl \,db \,d\varphi.
\end{aligned}
\label{eq:n_r_small}
\end{equation}

Function $p_6 $(Eq. ~\ref{eq:n_r,v,B,l,b,phi}) can be integrated analytically with respect to $\varphi$ and $B$, giving

\begin{equation}
\begin{aligned}
&p_{r, B_{cdf}, \varphi_{cdf}}=\\
&\begin{cases}
\begin{aligned}
\displaystyle
&\frac{\varphi}{4\pi} \int \limits_{v_{\infty\left(min\right)}}^{v_{\infty\left(max\right)}}\int \limits_{0}^{2\pi} \int \limits_{-\frac{\pi}{2}}^{\frac{\pi}{2}} p_{v_{\infty}, l, b}\\
&\left(\sqrt{1+\frac{2\mu}{r v_{\infty}^2}}-\sqrt{1+\frac{2\mu}{r v_{\infty}^2}-\frac{B^2}{r^2}}\right) \,dv_{\infty} \,dl \,db
\end{aligned}, & \text{$B \leq B_1$} \\\\\\
\begin{aligned}
\displaystyle
&\frac{\varphi}{4\pi} \int \limits_{v_{\infty\left(min\right)}}^{v_{\infty\left(max\right)}}\int \limits_{0}^{2\pi} \int \limits_{-\frac{\pi}{2}}^{\frac{\pi}{2}} p_{v_{\infty}, l, b}\\ &\left(\sqrt{1+\frac{2\mu}{r v_{\infty}^2}}\right.\\
\displaystyle
&\left.+\sqrt{1+\frac{2\mu}{r v_{\infty}^2}-\left(\frac{R}{r}\right)^2\left(1+\frac{2\mu}{v_{\infty}^2R}\right)}\right.\\ 
\displaystyle
&\left.-2\sqrt{1+\frac{2\mu}{r v_{\infty}^2}-\frac{B^2}{r^2}}\right)\,dv_{\infty} \,dl \,db
\end{aligned}, & \text{$B_1< B \leq B_2$}
\end{cases}
\end{aligned}.
\label{eq:p_r,B_cdf,phi_cdf}
\end{equation}

As stated above, this is a probability density function of $r$ that is cumulative with respect to $B$ and $\varphi$ and marginal with respect to $v_{\infty}$, $l$ and $b$ (i.e. integrated over the whole domain). In marginal cases also for $B$ and $\varphi$, when $B=B_1$ (for impactors) or $B=B_2$ (for both impactors and passers) and $\varphi=2\pi$, function $p_{r, B_{cdf} , \varphi_{cdf}}$ becomes marginal with respect to all parameters, except $r$, in the form

\begin{equation}
\begin{aligned}
&p_r=\\
&\begin{cases}
\begin{aligned}
\displaystyle
&\int \limits_{v_{\infty\left(min\right)}}^{v_{\infty\left(max\right)}}\int \limits_{0}^{2\pi} \int \limits_{-\frac{\pi}{2}}^{\frac{\pi}{2}} \frac{p_{v_{\infty}, l, b}}{2}\left(\sqrt{1+\frac{2\mu}{r v_{\infty}^2}}\right.\\
\displaystyle
&\left.-\sqrt{1+\frac{2\mu}{r v_{\infty}^2}-\left(\frac{R}{r}\right)^2\left(1+\frac{2\mu}{v_{\infty}^2 R}\right)}\right)\,dv_{\infty} \,dl \,db
\end{aligned}, & \text{$B=B_1$} \\\\\\
\begin{aligned}
\displaystyle
&\int \limits_{v_{\infty\left(min\right)}}^{v_{\infty\left(max\right)}}\int \limits_{0}^{2\pi} \int \limits_{-\frac{\pi}{2}}^{\frac{\pi}{2}}\frac{p_{v_{\infty}, l, b}}{2}\left(\sqrt{1+\frac{2\mu}{r v_{\infty}^2}}\right.\\
\displaystyle
&\left.+\sqrt{1+\frac{2\mu}{r v_{\infty}^2}-\left(\frac{R}{r}\right)^2\left(1+\frac{2\mu}{v_{\infty}^2 R}\right)}\right)\,dv_{\infty} \,dl \,db
\end{aligned}, & \text{$B=B_2$}
\end{cases}
\end{aligned}.
\label{eq:p_r}
\end{equation}

\begin{figure}[h]
	\includegraphics[width=\columnwidth]{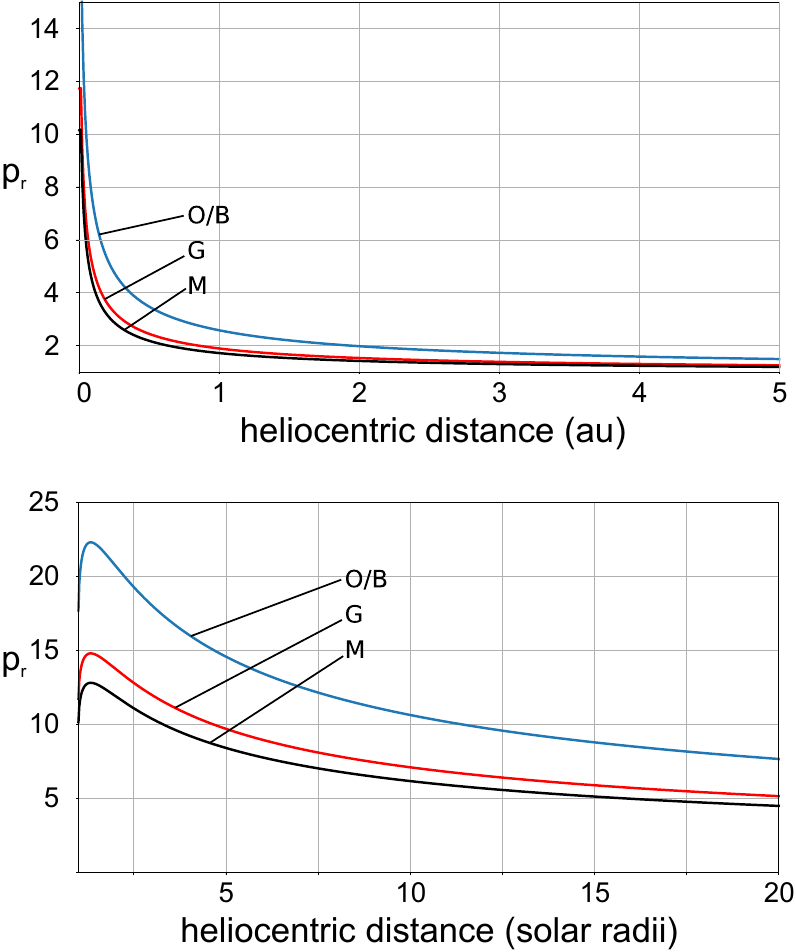}
    \caption{Variation of number-density of ISOs with the heliocentric distance, assuming kinematics of stellar classes M, G and O/B (see Table ~\ref{table:2}). The top panel shows this variation over the larger range of heliocentric distances that traverse the planetary orbits. The bottom panel shows this variation at small heliocentric distances comparable to the solar radius.}
    \label{fig:p_r}
\end{figure}

Integration in Eq. ~\ref{eq:p_r} with respect to $v_{\infty}$, $l$ and $b$ is conducted numerically to obtain variation of number-density with heliocentric distance. All results presented in this paper were obtained by applying Simpson's integration rule with steps $\Delta r = 0.1$ au, $\Delta v = 1$ km/s, $\Delta l=\Delta b = 1^{\circ}$. The variations of number-density with heliocentric distance for the three velocity distributions defined in Table ~\ref{table:2} are shown in Figure ~\ref{fig:p_r}.

Objects that approach very close to the Sun may perish before the perihelion passage due to disintegration, outgassing torques, tidal disruption, etc. In order to survive, they must satisfy the so-called \emph{perihelion survival limit} \citet{1991ICQ....13...89B}. Observational constraints on the behavior of sungrazing comets near perihelion come mostly from the well known Kreutz group. These objects do not survive perihelion. Their brightness usually peaks at 10–15 solar radii and then rapidly fade \citep{2013ApJ...776L...5K}. In the context of the \emph{Probabilistic method}, these objects should be viewed as solar impactors because they contribute to the population only on the inbound branches of their orbits. In this sense, the radius $R$ in Eqs. ~\ref{eq:B_lim}, ~\ref{eq:p_r,B_cdf,phi_cdf} and ~\ref{eq:p_r} should be considered as the effective radius that protects the space behind from incoming objects. Obviously, this radius depends on the physical characteristics of the object, its rotational state, etc, and it can be defined as input parameter in the \emph{Probabilistic method}. This effect does not change the character of the curves shown in the bottom panel of Fig. ~\ref{fig:p_r}, but shifts them to greater heliocentric distances.

The total number of objects inside an infinitesimal spherical shell with an arbitrary radius $r$ is

\begin{equation}
N_{dr}=4\pi r^2 p_r n_0 dr.
\label{eq:p_r,v}
\end{equation}

The total number of objects inside a heliocentric sphere with radius $r$ is obtained by integrating $N_{dr}$ with respect to $r$, 

\begin{equation}
N_r=4\pi n_0\int_{R}^{r}{p_{r} r^2 dr},
\label{eq:N_r}
\end{equation}
where $R$ is the effective radius of the Sun.

The 6-step procedure for sampling the 6 parameters used to determine orbital elements of ISOs works as follows.

\subsection*{Step 1}

\begin{figure}[h]
	\includegraphics[width=\columnwidth]{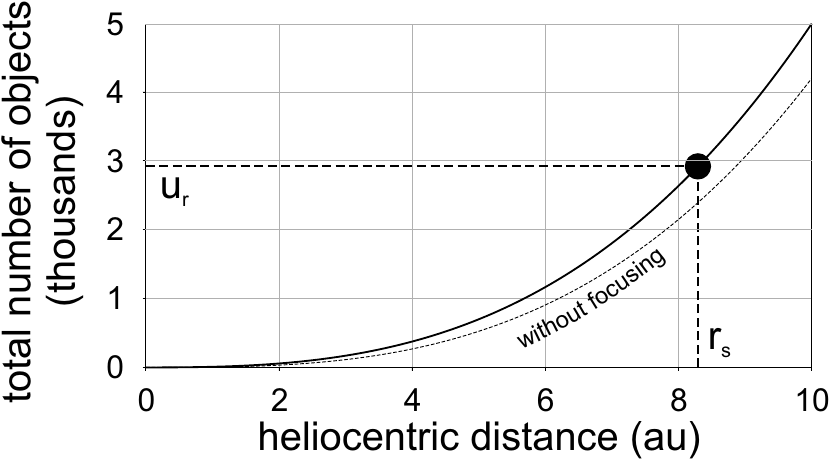}
    \caption{Total number of objects in the sphere around the Sun. The upper curve shows how the total number of ISOs increases if the radius of heliocentric sphere increases, assuming velocity distribution of interstellar velocities of M-stars(see Table ~\ref{table:2}) and interstellar number-density $n_0=1$. The curve representing corresponding number of objects without gravitational focusing is also given as a reference. This curve actually presents the variation of the volume of sphere with its radius. A schematic illustration of determining the heliocentric distance of a synthetic object ($r_s$) by using the inverse transform sampling method is also shown.}
    \label{fig:N_r}
\end{figure}

Function $N_r$ allows sampling of heliocentric distances of ISOs using the inverse transform sampling method. As schematically illustrated in Fig. ~\ref{fig:N_r}, a uniform sample $\boldsymbol{u_r}=U\left(0, max\left(N_r\right)\right)$ is transformed into a set of heliocentric distances $\boldsymbol{r_s}$, 

\begin{equation}
\boldsymbol{r_s}=N_r^{-1}\left(\boldsymbol{u_r}\right).
\label{eq:r_s}
\end{equation}

The size of the set is equal to the total number of objects, i.e. an integer value of $max\left(N_r\right)$. Function $N_r$ is obtained by numerical integration and it is in a discrete form, so the inversion has to be conducted by using some of the methods of inverse interpolation. All results presented in this paper were obtained by using the cubic spline interpolation. Further decreasing of the integration steps does not change results significantly. The obtained results also do not depend significantly on the choice of the interpolation method, especially when the integration steps in Eqs. ~\ref{eq:n_r_small}, ~\ref{eq:N_r}, ~\ref{eq:p_v_cdf(r_s)}, ~\ref{eq:p_B_cdf}, ~\ref{eq:p_l_cdf} and ~\ref{eq:p_lat_cdf} are small. The impact of the interpolation procedure is addressed in Section ~\ref{Results 1}.

\subsection*{Step 2}

Once the set of heliocentric distances ${\bf r_s}$ is generated, the next step is to generate the corresponding set of interstellar velocities ${\bf v_{\infty s}}$. As pointed out earlier, this is done by inverting a probability function which is conditional with respect to already determined heliocentric distances, cumulative with respect to $v_{\infty}$, and marginal with respect to other yet to be determined parameters. In order to sample from uniform random number $U\left(0,1\right)$, the function has to be normalized to $p_r\left(r_s\right)$. For each element $r_s$ from set ${\bf r_s}$ this function is obtained by integration

\begin{equation}
\begin{aligned}
p_{v_{\infty cdf}}\left(r_s\right)=&\frac{1}{p_r\left(r_s\right)}\int \limits_{v_{\infty \left(min\right)}}^{v_{\infty}}\int \limits_{0}^{B_2}\int \limits_{0}^{2\pi}\int \limits_{-\frac{\pi}{2}}^{\frac{\pi}{2}} \int \limits_{0}^{2\pi}\\&p_6\left(r_s \right)\,dv_{\infty} \,dB \,dl  \,db \,d\varphi.
\end{aligned}
\label{eq:p_v_cdf(r_s)}
\end{equation}

Interstellar velocities $v_{\infty s}$ that correspond to every $r_s$ from set $\bf r_s$ are obtained by inverting this function,

\begin{equation*}
v_{\infty s}=p_{v_{\infty cdf}} \left(r_s\right)^{-1}\left(u_v\right),\quad u_v\sim U\left(0,1\right).
\end{equation*}

\begin{figure}[h]
	\includegraphics[width=\columnwidth]{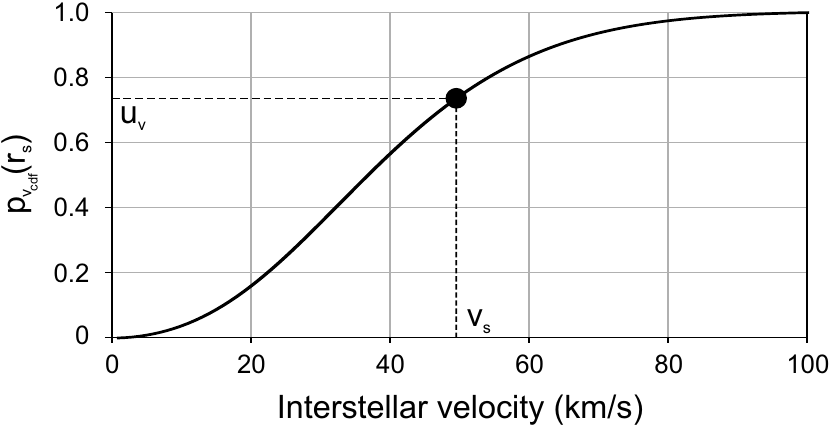}
    \caption{The cumulative distribution function of interstellar velocities for objects with defined heliocentric distance $r_s$, which is marginal with respect to impact parameters and longitudes and latitudes of the interstellar velocity vector. Schematic illustration of inverting this function to obtain $v_{\infty s}$ is also shown.}
    \label{fig:v_s}
\end{figure}

As shown in Eq. ~\ref{eq:p_r,B_cdf,phi_cdf}, this function can be integrated analytically with respect to $B$ and $\varphi$, while the integration with respect to $v_{\infty}$, $l$ and $b$ is conducted numerically. Fig. ~\ref{fig:v_s} shows this function and schematic illustration of its inversion to obtain $v_{\infty s}$.

\subsection*{Step 3}

After the sets of heliocentric distances ${\bf r_s}$ and corresponding interstellar velocities ${\bf v_{\infty s}}$ are determined, the impact parameters can be determined in an analogous manner. For each pair $\left(r_s, v_{\infty s}\right)$ from these sets, corresponding $B_s$ can be sampled from a probability function that is cumulative with respect to $B$, conditional with respect to $r_s$ and $v_{\infty s}$, and marginal with respect to other parameters $l$, $b$ and $\varphi$. Similar to the previous step, this function can be normalized to $p_{r,v_{\infty}}\left(r_s, v_{\infty s}\right)$, giving

\begin{equation}
\begin{aligned}
&p_{B_{cdf}} \left(r_s, v_{\infty s}\right)=\frac{1}{p_{r, v_{\infty}}\left(r_s, v_{\infty s}\right)} \\& \int \limits_{0}^{B}\int \limits_{0}^{2\pi}\int \limits_{-\frac{\pi}{2}}^{\frac{\pi}{2}} \int \limits_{0}^{2\pi}p_6\left(r_s, v_{\infty s}\right) \,dB \,dl  \,db \,d\varphi.
\end{aligned}
\label{eq:p_B_cdf}
\end{equation}

The corresponding impact parameter is obtained by inverting this function,

\begin{equation*}
B_s=p_{B_{cdf}} \left(r_s, v_{\infty s}\right)^{-1}\left(u_B\right),\quad u_B\sim U\left(0,1\right).
\end{equation*}

It should be noticed that function $p_{B_{cdf}}$ has two forms depending on whether the object is the Sun-impactor or not (see Eq. ~\ref{eq:p_r}). Which of these two forms is used is defined by the value of the random number $u_B$. If $u_B \leq p_{B_{cdf}}\left(r_s, v_{\infty s}, B_1\right)$ then the object is the Sun-impactor and the first form of Eq. ~\ref{eq:p_r,B_cdf,phi_cdf} is used. Otherwise, the object passes by the Sun and the second form is used (see also Eq. ~\ref{eq:B_lim} and Fig. ~\ref{fig:limit}). In both cases, unlike inverting functions $N_r$ and $p_{v_{\infty cdf}}$ to obtain $r_s$ and $v_{\infty s}$, function $p_{B_{cdf}} \left(r_s, v_{\infty s}\right)$ can be inverted analytically giving the resulting impact parameter $B_s$ in the form

\begin{equation}
\begin{aligned}
&B_s= \\
&\begin{cases}
\displaystyle
r_s\sqrt{f_1^2-\left(\frac{f_1 p_v\left(v_{\infty s}\right)-2u_B}{p_v\left(v_{\infty s}\right)}\right)^2} & \text{if $u_B\leq p_{B_{cdf}}\left(r_s, v_{\infty s}, B_1\right)$}\\\\\\
\displaystyle
r_s\sqrt{f_1^2-\left(\frac{f_2 p_v\left(v_{\infty s}\right)-2u_B}{2p_v\left(v_{\infty s}\right)}\right)^2} & \text{if $u_B>p_{B_{cdf}}\left(r_s, v_{\infty s}, B_1\right)$}
\end{cases}
\end{aligned},
\label{eq:B_s}
\end{equation}
where
\begin{equation*}
\begin{aligned}
\displaystyle
&f_1=\sqrt{1+\frac{2\mu}{r_s v_{\infty s}^2}}, \\\\
\displaystyle
&f_2=\sqrt{1+\frac{2\mu}{r_s v_{\infty s}^2}}+\sqrt{1+\frac{2\mu}{r_s v_{\infty s}^2}-\left(\frac{R}{r_s}\right)^2\left(1+\frac{2\mu}{R v_{\infty s}^2}\right)}
\end{aligned}.
\end{equation*}

\begin{figure}[h]
	\includegraphics[width=\columnwidth]{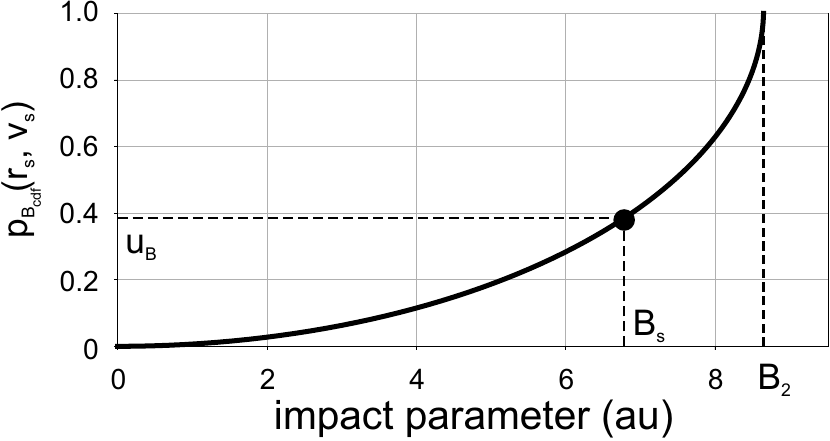}
    \caption{The cumulative distribution function of impact parameters, for objects with defined heliocentric distance $r_s$ and interstellar velocity $v_{\infty s}$, which is marginal with respect to longitudes and latitudes of the interstellar velocity vector. Schematic illustration of inverting this function to obtain $B_s$ is also shown.}
    \label{fig:B_s}
\end{figure}

Fig. ~\ref{fig:B_s} shows function $p_{B_{cdf}}$ and a schematic illustration of its inversion to obtain $B_s$.

\subsection*{Step 4}

The next step is to determine the longitude of the interstellar velocity vector for each corresponding group of already determined parameters ($r_s, v_{\infty s}, B_s$). This is obtained from a probability function that is cumulative with respect to $l$ , conditional with respect to $r_s$ , $v_{\infty s}$ and $B_s$, and marginal with respect to the remaining two still undetermined parameters $b$ and $\varphi$. Again, this function can be normalized to $p_{r,v_{\infty}, B}\left(r_s,v_{\infty s}, B_s \right)$, giving

\begin{equation}
\begin{aligned}
p_{l_{cdf}}\left(r_s, v_{\infty s}, B_s\right)=&\frac{1}{p_{r,v_{\infty}, B}\left(r_s,v_{\infty s}, B_s \right)} \int \limits_{0}^{l}\int \limits_{-\frac{\pi}{2}}^{\frac{\pi}{2}} \int \limits_{0}^{2\pi}\\&p_6\left(r_s,v_{\infty s}, B_s \right) \,dl  \,db \,d\varphi.
\end{aligned}
\label{eq:p_l_cdf}
\end{equation}

As in all previous step, this function can be integrated analytically with respect to $\varphi$, while the integration with respect to $l$ and $b$ has to be conducted numerically. Longitude of the interstellar velocity vector is obtained by inverting this function

\begin{equation*}
l_s=p_{l_{cdf}} \left(r_s, v_{\infty s}, B_s\right)^{-1}\left(u_l\right),\quad u_l\sim U\left(0,1\right).
\end{equation*}

\begin{figure}[h]
	\includegraphics[width=\columnwidth]{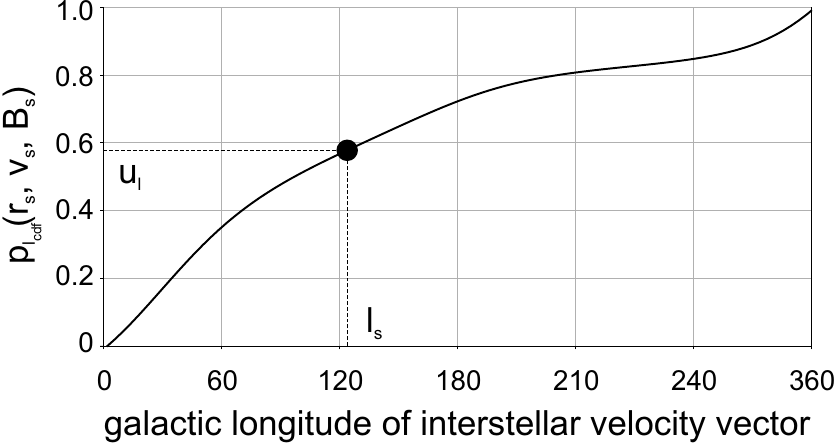}
    \caption{The cumulative distribution function of longitude of interstellar velocity vector, for objects with defined heliocentric distance $r_s$, interstellar velocity $v_{\infty s}$ and impact parameter $B_s$, which is marginal with respect to latitudes of the interstellar velocity vector. Schematic illustration of inverting this function to obtain $l_s$ is also shown.}
    \label{fig:l_s}
\end{figure}

Fig. ~\ref{fig:l_s} shows this function and schematic illustration of its inversion to obtain $l_s$.

\subsection*{Step 5}

The fifth step is to determine latitude of the interstellar velocity vector for every corresponding group of already determined parameters $\left(r_s, v_{\infty s}, B_s, l_s\right)$. This is obtained from a probability function which is cumulative with respect to $b$, conditional with respect to $r_s$, $v_{\infty s}$, $B_s$ and $l_s$ and marginal with respect to the remaining undetermined parameter $\varphi$. As in all previous steps, this function can be normalized to $p_{r,v_{\infty}, B, l}\left(r_s,v_{\infty s}, B_s, l_s \right)$, giving

\begin{equation}
\begin{aligned}
&p_{b_{cdf}}\left(r_s, v_{\infty s}, B_s, l_s\right)=\frac{1}{p_{r, v_{\infty}, B, l}\left(r_s, v_{\infty s}, B_s, l_s\right)}\\& \int \limits_{-\frac{\pi}{2}}^{b} \int \limits_{0}^{2\pi} p_6\left(r_s, v_{\infty s}, B_s, l_s,\right)  \,db \,d\varphi.
\end{aligned}
\label{eq:p_lat_cdf}
\end{equation}

This function can be integrated analytically with respect to $\varphi$ (see Eq. ~\ref{eq:p_B_cdf}), while the integration with respect to $b$ has to be conducted numerically. Latitude of the interstellar velocity vector is obtained by inverting this function

\begin{equation*}
b_s=p_{b_{cdf}} \left(r_s, v_{\infty s}, B_s, l_s\right)^{-1}\left(u_b\right),\quad u_b\sim U\left(0,1\right).
\end{equation*}

\begin{figure}[h]
	\includegraphics[width=\columnwidth]{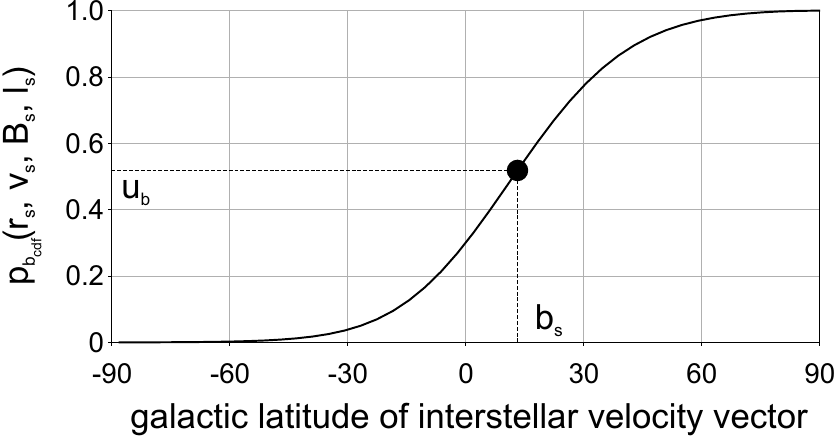}
    \caption{The cumulative distribution function of latitude of interstellar velocity vector, for objects with defined heliocentric distance $r_s$, interstellar velocity $v_{\infty s}$, impact parameter $B_s$ and longitude $l_s$. Schematic illustration of inverting this function to obtain $b_s$ is also shown.}
    \label{fig:small_b_s}
\end{figure}

Fig. ~\ref{fig:small_b_s} shows this function and schematic illustration of its inversion to obtain $b_s$.

\subsection*{Step 6}

The last step is to determine angle $\varphi$. As mentioned earlier, the joint probability function $p_6$ does not depend on this angle, which means that it can simply be sampled from a uniform distribution $\varphi \sim U\left(0, 2\pi\right)$.

\section{Results and discussion}
\label{Results and discussion}

The \emph{Probabilistic method} is applied to generate synthetic populations of ISOs assuming different distributions of interstellar. The resulting populations are compared to those generated by the \emph{Dynamical methods} and numerical integration in order to analyse its ability to generate realistic populations and its computational performances.

\subsection{Sensitivity of the ISO population on the assumed distribution of interstellar velocities}
\label{Results 1}

\begin{figure}[h]
	\includegraphics[width=\columnwidth]{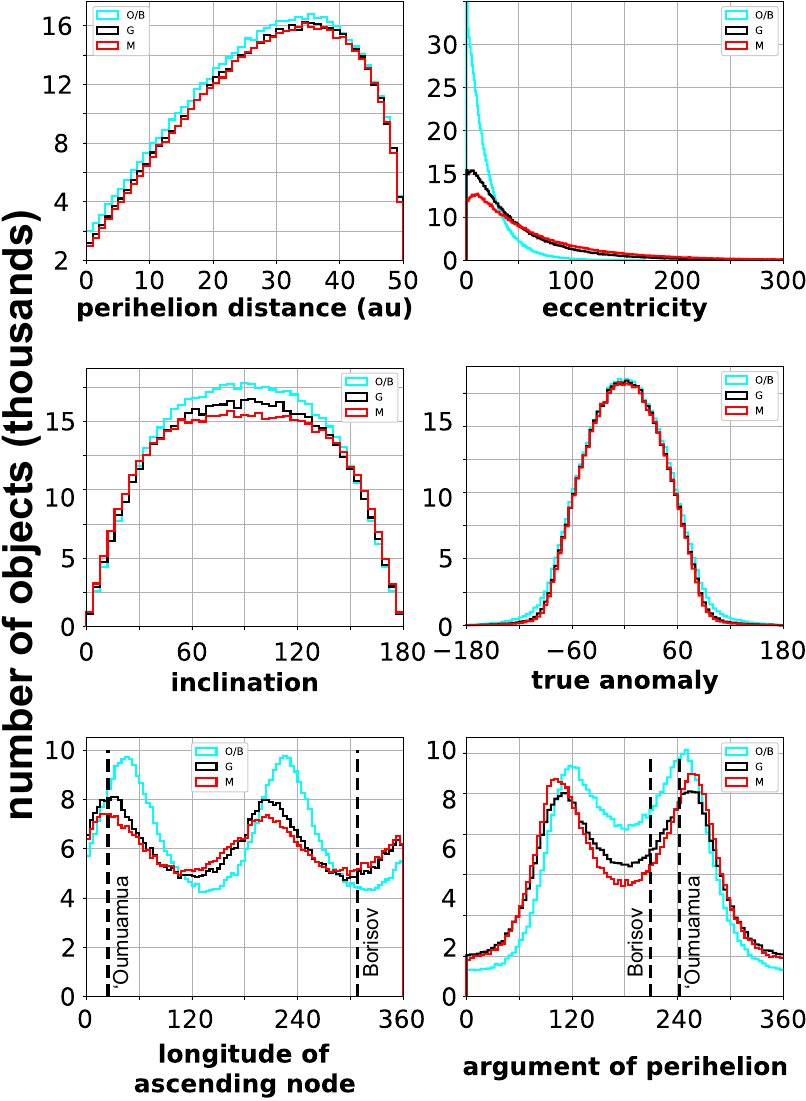}
    \caption{Distributions of orbital elements of ISOs within a heliocentric sphere of radius 50 au, assuming kinematics of stellar classes M, G and O/B (see Table ~\ref{table:2}). For reference, vertical lined denoting 1I/'Oumuamua and 2I/Borisov are given on the plots for argument of perihelion and longitude of ascending node, the two orbital elements likely least affected by observational selection effects.}
    \label{fig:orbital elements}
\end{figure}

To demonstrate the sensitivity of the ISO population to the assumed distribution of interstellar velocities, the \emph{Probabilistic method} was used to generate synthetic populations of ISOs within a heliocentric sphere of radius 50 au, for three different distributions of interstellar velocities defined in Table ~\ref{table:2}. Histograms of 6 orbital elements of the obtained synthetic populations are shown in Fig. ~\ref{fig:orbital elements}.

Fig. ~\ref{fig:orbital elements} shows that the O/B-stars kinematics, due to the smaller velocity component dispersions, results in the largest number of objects, which in this case is 577177, compared to 555009 and 542266 for the G- and M-stars kinematics, respectively. This can be seen most clearly in histograms of perihelion distances, orbital inclinations and true anomalies. In addition, there are noticeable differences in the distributions of eccentricity, argument of perihelion, and longitude of ascending node. The orbits of the population with the kinematics of O/B stars are distributed over a much smaller range of eccentricities. The mean eccentricity values are around 20, 49 and 64 for the populations with O/B, G and M-star kinematics, respectively. In addition, there are noticeable differences in the distributions of the argument of perihelion $\omega$ and longitude of ascending node $\Omega$, especially for the population with O$/$B kinematics which has more pronounced variations. The velocity dispersions of the kinematics of O/B-stars are the smallest and consequently, the corresponding distributions of $\omega$ and $\Omega$ are most affected by the solar motion with respect to the LSR. Consequently, the objects that originate from these stars are probably more likely to be captured by the solar system.

\begin{figure}[h]
	\includegraphics[width=\columnwidth]{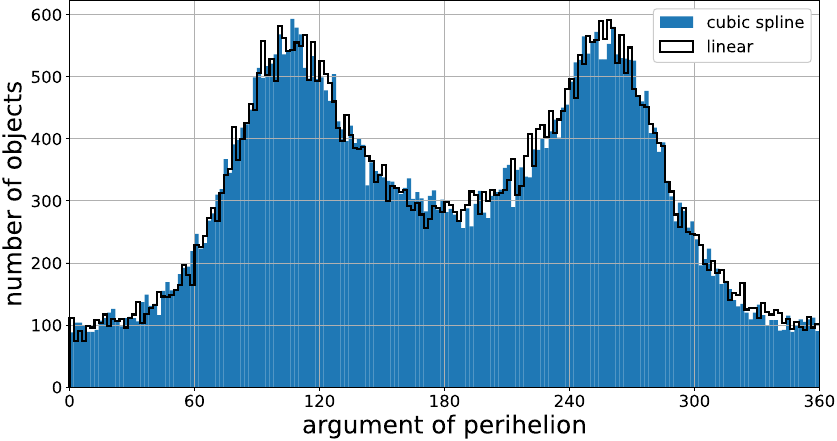}
    \caption{Comparison of distributions of argument of perihelion obtained using cubic spline and linear interpolation.}
    \label{fig:interpolation}
\end{figure}

As mentioned earlier, the results reported in Fig. ~\ref{fig:orbital elements} were obtained using cubic spline interpolation. The \emph{Probabilistic method} can be easily adjusted to use any other interpolation method. However, it should not lead to significant change of the results. For example, Fig. ~\ref{fig:interpolation} shows comparison of distributions of argument of perihelion using cubic spline and simple linear interpolation. Although the steps in this case were much larger ($\Delta v=2$ km/s, $\Delta r=1$ au, $\Delta l=\Delta b=4^{\circ}$), the obtained results are very consistent since cosine similarity between the two histograms shown in Fig. ~\ref{fig:interpolation} is $\sim$ 0.999.

\begin{figure}[h]
	\includegraphics[width=\columnwidth]{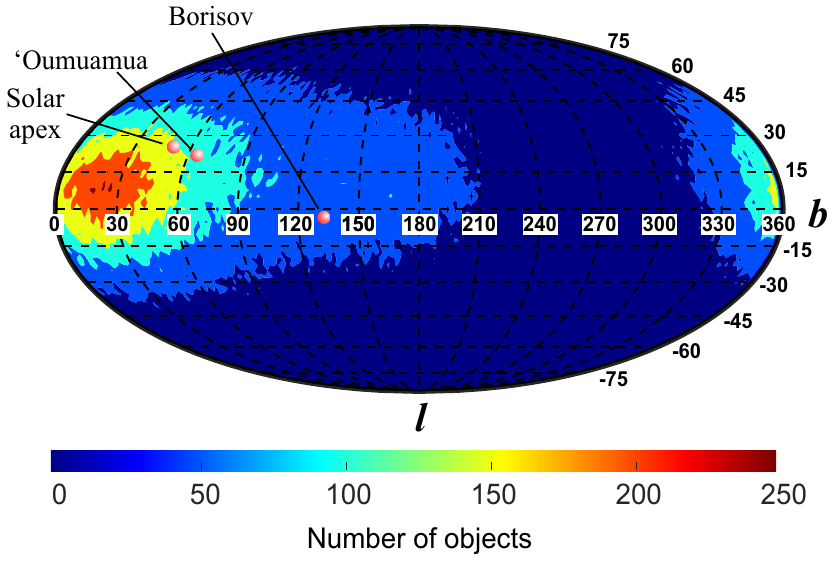}
    \caption{Distribution of incoming directions of ISOs in the galactic reference frame for the synthetic population inside model sphere with radius of 50 au, number-density of 1 object per au$^3$ and kinematics of M-stars (table ~\ref{table:2})}
    \label{fig:sky_map}
\end{figure}

Fig. ~\ref{fig:sky_map} shows distribution of the incoming directions of ISOs, assuming the kinematics of M-stars (see Table ~\ref{table:2}). The majority of objects come from the region near the solar apex. The densest region is slightly shifted toward the galactic center due to the greater dispersion of U-components of the interstellar velocity vectors.

The comparison of these three cases shows how important it is to have an adequate assumption about the interstellar velocity distribution of ISOs in order to synthesize their realistic population. The interstellar kinematics of ISOs may depend not only on the global kinematics of the stars in the solar neighborhood, but also on the relative contribution of different stellar classes to this population. In this way, the ISO kinematics may result from a specific mixture of stellar kinematics. When a sufficient number of ISOs are discovered to reconstruct their orbital structure in the Solar System, it could give an indication of the contribution of different stellar classes to the galactic population of these objects. As it is shown in Section \ref{Results 2}, the \emph{Probabilistic method} is very computationally efficient, so it can be used to generate a large number of synthetic populations with a variety of input kinematics. In this way, one can infer which mixture resulted in the actual population, thus indicating the relative contributions of the different stellar classes to the ISO population.

\subsection{Comparison of Probabilistic and Dynamical methods}
\label{Results 2}

\begin{figure}[h]
	\includegraphics[width=\columnwidth]{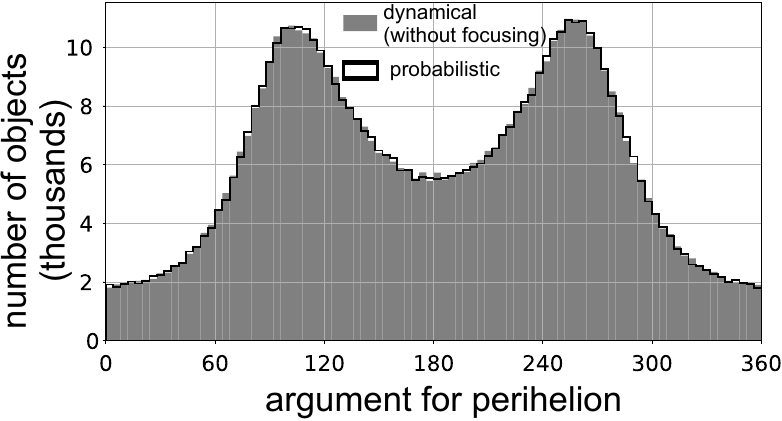}
    \caption{Histograms of argument of perihelion obtained by the \emph{Probabilistic method} and \emph{Dynamical method (without focusing)}.}
    \label{fig:dyn_vs_prob}
\end{figure}

In order to cross-check the results obtained by the \emph{Dynamical} and \emph{Probabilistic} methods, a comparison was made for the populations generated within the model sphere with a radius of 50 au, number-density of 1 object per au$^3$ and the kinematics of M-stars (see table ~\ref{table:2}). The obtained distributions of orbital elements are practically indistinguishable. To quantify the consistency of the results, the cosine similarities of the corresponding histograms were calculated and none were below 0.999, regardless of the bin size. For visual comparison, histograms of argument of perihelion is shown in Fig. ~\ref{fig:dyn_vs_prob}.

As mentioned earlier, there are two main advantages of the \emph{Probabilistic} over the \emph{Dynamical methods}. One refers to accuracy of the synthesized population, i.e. how well it corresponds to the input distribution of interstellar velocities, while the other one refers to computational efficiency.

\begin{figure}[h]
	\includegraphics[width=\columnwidth]{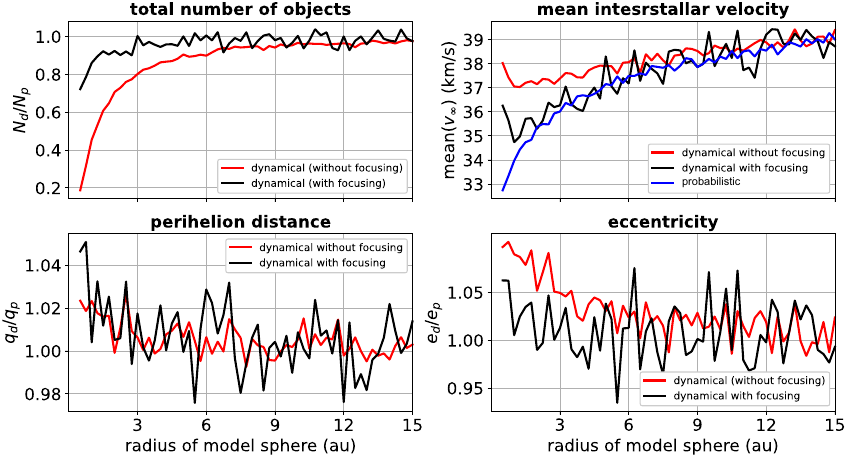}
    \caption{Comparison of parameters of populations generated using the \emph{Probabilistic} and \emph{Dynamical methods}.}
    \label{fig:parameter_comparison}
\end{figure}

The \emph{Dynamical method (without focusing)} is accurate enough only if it is used to generate a synthetic population in a sufficiently large heliocentric sphere. If the population is generated in a small heliocentric sphere, it underestimates the total number of objects, and the population is also biased toward faster objects. This means that even a realistic kinematics of ISOs in interstellar space is provided, the resulting synthetic population will deviate systematically from the real population. As previously discussed, this is due to the fact that the \emph{Dynamical method (without focusing)} assumes that outside the sphere where the population is generated, the gravitational focusing is neglected and the number-density is equal to that in interstellar space. However, as illustrated in Fig. ~\ref{fig:p_r}, this effect affects the population at large heliocentric distances and its neglect leads to certain underestimation of the number of objects in the inner Solar System. On the other hand, \emph{Dynamical method (with focusing)} takes this effect into account, but leaves the directions of the initial velocity vectors unaffected by the solar gravity.
To quantify the influence of these effects, synthetic populations were generated using the \emph{Dynamical} and \emph{Probabilistic methods} within the model spheres of radii from 0.5 au to 15 au, with a step of 0.25 au. The populations were generated assuming velocity distribution of M-stars (see Table ~\ref{table:2}) and the results are reported in Fig. ~\ref{fig:parameter_comparison}.

Fig. ~\ref{fig:parameter_comparison} (top-left panel) shows, that the \emph{Dynamical method (without focusing)} significantly underestimates the total number of objects for populations generated in small model spheres, but this number closely approaches that of the \emph{Probabilistic method} for populations created within a heliocentric sphere with a radius greater than about 10 au. On the other hand, \emph{Dynamical method (without focusing)} shows much better agreement with the \emph{Probabilistic method}, resulting in practically the same number of objects for model spheres with radii larger than 3 au.

As previously discussed, gravitational focusing affects stronger objects with lower interstellar velocities and ignoring it leads not only to underestimating the total number of objects, but also to changing the population structure by reducing the relative participation of slower objects. Fig. ~\ref{fig:parameter_comparison} (top-right panel) shows mean interstellar velocities for populations generated using the \emph{Probabilistic} and \emph{Dynamical} methods in model spheres with radii ranging from 0.5 to 15 au. In this case, \emph{Dynamical method (with focusing)} shows excellent agreement with the \emph{Probabilistic method}, while the \emph{Dynamical method (without focusing)} has larger values at small heliocentric distances due to the bias toward faster objects. Underestimation of slower objects is significant because they are very important (e.g. they have the highest probability to be captured by some of the dynamical mechanisms \citep{2021MNRAS.tmp.3398D, 2021PSJ.....2...53N, 2021PSJ.....2..217N}).

The bottom two panels in Fig. ~\ref{fig:parameter_comparison} show deviation of mean perihelion distance and mean eccentricity for populations generated by the \emph{Dynamical methods} from the one generated using the \emph{Probabilistic method}. Again, as expected, \emph{Dynamical method (with focusing)} has much better agreement. However, as mentioned earlier, although this method accounts for the gravitational focusing outside the model sphere, and adjusts the initial speeds for the solar acceleration, it ignores the fact that the directions of the velocity vectors at an arbitrary heliocentric distance are not identical to the corresponding directions at infinity. This gives the orbits a slightly larger angular momentum, and therefore a slightly higher eccentricity and perihelion distance.

Fig. ~\ref{fig:parameter_comparison} shows that populations generated using the \emph{Dynamical method} converge to those generated using the \emph{Probabilistic method} if the heliocentric distance increases. As a consequence, if a population of ISOs is generated using the \emph{Dynamical methods}, it has to be generated within a large model sphere, beyond which the effect of gravitational focusing can be safely ignored. For example, if one needs a synthetic population of ISOs around the Earth's orbit, the population has to be generated in a much larger volume of space and then the relevant objects from that population have to be selected.

\begin{figure}[h]
	\includegraphics[width=\columnwidth]{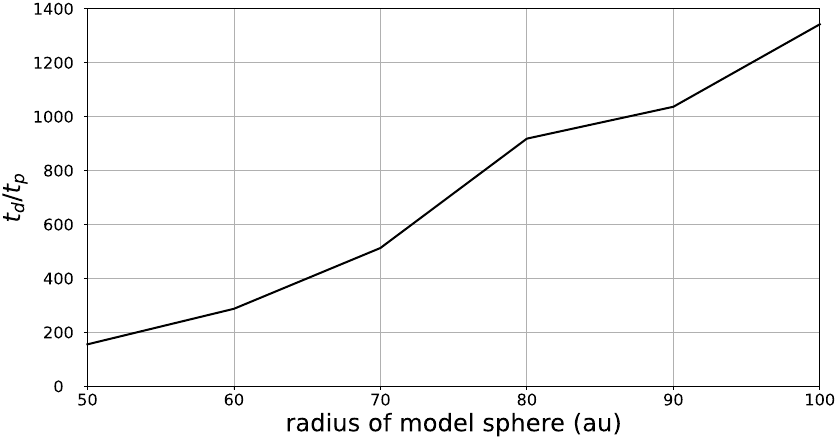}
    \caption{Comparison of the times required to generate synthetic populations using \emph{Dynamical method (with focusing)} ($t_d$) and \emph{Probabilistic method} ($t_s$). Due to the very large computational demand of the \emph{Dynamical method}, these data were obtained by generating populations assuming very small number-density $n_0=10^{-3}$ object per au$^3$.}
    \label{fig:time_ratio}
\end{figure}

Another advantage of the \emph{Probabilistic method} relates to its computational efficiency, since it is significantly faster than the \emph{Dynamical method}. Fig. ~\ref{fig:time_ratio} shows the ratio of the times required to generate synthetic populations by \emph{Dynamical method} ($t_d$) and \emph{Probabilistic method} ($t_s$), depending on the radius of the model sphere.

As Fig. ~\ref{fig:time_ratio} illustrates, this ratio increases with the increased volume of space where the synthetic population is generated, reaching three orders of magnitude for a model sphere with radius of 100 au. Analysing ISO populations usually requires synthetic populations within such large heliocentric spheres. For example, \citet{Engelhardt2017} and \citet{2020MNRAS.498.5386M} analysed synthetic populations inside model spheres with radii 50 au and 100 au, respectively. The reason for this was that objects within these spheres are capable of reaching heliocentric distances at which they could be detected by the LSST survey during the defined nominal duration of this survey of 10 years \citep{2017arXiv170804058L}. With this in mind, the computational efficiency of the procedure becomes very important for the possibility of its systematic implementation. The computational efficiency of the \emph{Probabilistic method} depends on the choice of the integration steps in Eqs. ~\ref{eq:n_r_small}, ~\ref{eq:N_r}, ~\ref{eq:p_v_cdf(r_s)}, ~\ref{eq:p_B_cdf}, ~\ref{eq:p_l_cdf} and ~\ref{eq:p_lat_cdf}. Steps $dr=10^{-3}$ au, $dv=1$ km/s, $dl=db=1^{\circ}$ were used for the comparison reported in Fig. ~\ref{fig:time_ratio}. Further decreasing of the integration steps does not lead to a significant change of the results, as previously discussed in Section \ref{Probabilistic method}. On the other hand, most of the computational load in the \emph{Dynamical methods} goes to the generation of initial Cartesian state vectors and their conversion to orbital elements and solving the hyperbolic Kepler equation for a large number of the initial objects. Adjusting the tolerance for numerical solving of this equation can change the computational efficiency of the method to a lesser extent, since the numerical schemes for solving the hyperbolic Kepler equation converge very quickly.

\subsection{Limitations of the Probabilistic method}
\label{Results 3}

The \emph{Probabilistic method} is completely heliocentric, which means that it takes into account only the gravitational influence of the Sun on ISOs and ignores the influence of the planets. However, when an ISO enters the Solar System it is initially governed by barycentric attraction. Analyzes of long-period comet orbits show that this can be good approximation up to heliocentric distances of about 150-200 au \citep{1981AcA....31..191T, 2017A&A...604A..24F}. In order to accurately track the orbit of the object at smaller heliocentric distances, perturbational effect of the planets has to be taken into account, as well as non-gravitational effects for comets at some point. In addition, close approach to the giant planets, especially Jupiter, can significantly change the orbital elements of the object \citep[e.g.][]{2013RMxAA..49..111B, 1982M&P....26..311B}. However, although these effects can have significant impact on a particular orbit, their impact on the orbital structure of ISOs in the Solar System is negligible. To demonstrate this, a population of ISOs is generated with the \emph{Probabilistic method} and compared to the population generated by numerical integration using Rebound ias15 numerical integrator \citep{2015MNRAS.446.1424R}, with all 8 planets included in the dynamical model. To obtain the population using numerical integration, the initial positions and velocity vectors were obtained in the same way as in the \emph{Dynamical method (without focusing)}. After that, the orbits were propagated by numerical integration that takes into account the full dynamical model, instead of solving the hyperbolic Kepler equation for each object, as in the \emph{Dynamical Model}. The populations were generated assuming M-stars kinematics (table ~\ref{table:2}).

\begin{figure}[h]
	\includegraphics[width=\columnwidth]{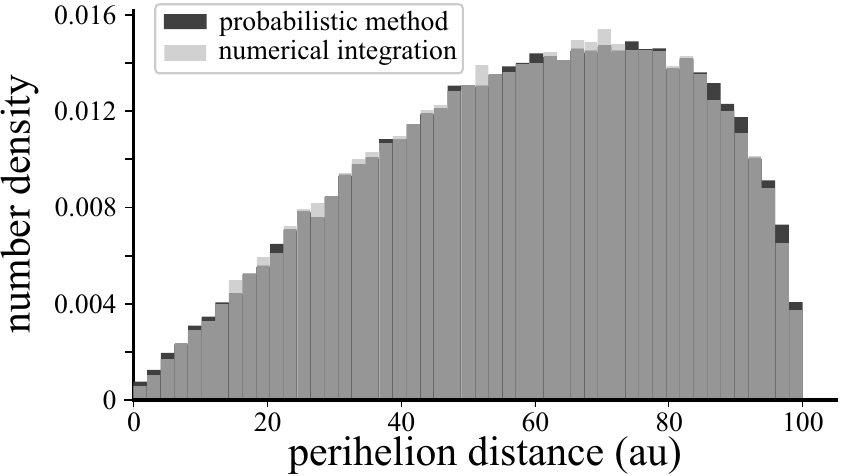}
    \caption{Normalised histograms of perihelion distance obtained by the \emph{Probabilistic method} and numerical integration.}
    \label{fig:statistical_vs_integration}
\end{figure}

Fig. ~\ref{fig:statistical_vs_integration}  shows the resulting histograms of perihelion distance obtained by the \emph{Probabilistic method} and numerical integration.

As Fig. ~\ref{fig:statistical_vs_integration} illustrates, the distributions of perihelion distance obtained by the \emph{Probabilistic method} and numerical integration are practically indistinguishable. The cosine similarity for these two histograms is $\sim 0.999$. This shows that planetary perturbations, although capable of changing a particular orbit significantly, cannot change the probability distributions of the orbital elements, so their impact on the orbital structure of instantaneous population of ISOs is negligible.

\section{Summary and Conclusions}
\label{Summary and Conclusions}

The 6-variate distribution of parameters defining orbits ISOs in the Solar System is derived. This distribution is used as a basis for the development of the so-called \emph{Probabilistic method} for generating the synthetic populations of ISOs.

The main conclusions can be summarised as follows:

\begin{enumerate}

    \item The effect of gravitational focusing defines the orbital structure of the population of interstellar objects around the Sun. The intensity of this effect depends on the assumed distribution of interstellar velocities of these objects. For the reasonable assumptions of these distributions, it increases number-density of ISOs by about factor of 2 around the Earth's orbit, and by order of magnitude at very small heliocentric distances.
    
    \item At very small heliocentric distances comparable to the Solar radius, the physical size of the Sun causes the so-called shielding effect, which results in a decrease of ISOs number-density. This space is populated only with interstellar Sun-impactors.
    
    \item The developed \emph{Probabilistic method} for generating synthetic populations of ISOs fully takes into account the effect of gravitational focusing, regardless of the radius of the sphere where the synthetic population is generated. This characteristic makes it suitable for generating synthetic population in small model spheres. In contrast, the \emph{Dynamical methods} underestimates the total number of objects when used to generate a synthetic population in small heliocentric spheres.
    
    \item The \emph{Dynamical methods} underestimates the number of slower objects in the populations. This shortcoming is significantly overcome by taking into account the gravitational focusing outside the sphere in which the synthetic population is generated. To entirely overcome this issue, the population using the \emph{Dynamical methods} has to be generated in a sufficiently large model sphere beyond which the effect of gravitational focusing can be safely ignored. On the other hand, the \emph{Probabilistic method} is independent of the size of the model sphere.
    
    \item The \emph{Probabilistic method} is significantly more efficient than the \emph{Dynamical methods}. The difference in efficiency depends on the size of the model sphere where the synthetic population is generated and reaches three orders of magnitude for a model sphere with radius of 100 au, depending on the numerical parameters involved in these two methods. High efficiency of the \emph{Probabilistic method} allows its systematic use for exploration of possible populations of ISOs.
    
\end{enumerate}

\section*{Data Availability Statement}
The Python code that implements the developed \emph{Probabilistic method} for generating the synthetic population of ISOs is available within \emph{Astronomy Source Code Library ASCL} at https://ascl.net/2209.014. The code includes Python libraries \emph{NumPy} \citep{harris2020array}, \emph{SciPy} \citep{2020SciPy-NMeth}, \emph{Random} \citep{van1995python} and \emph{tqdm} \citep{da_Costa-Luis2019}.

\begin{comment}
\section*{Acknowledgement}
Author acknowledges support by the Ministry of Education, Science and Technological Development of the Republic of Serbia, contract No. 451-03-68/2022-14/200104.
\end{comment}

\bibliographystyle{elsarticle-harv} 
\bibliography{cas-refs}





% \end{thebibliography}
\end{document}